\documentclass[traditabstract,twocolumn]{aa}

\usepackage{epsfig,graphicx,natbib}
\usepackage{longtable}
\usepackage{amssymb,amsmath}

\def\Tcmb{\hbox{$T_\mathrm{CMB}$}}
\def\kms{\hbox{km\,s$^{-1}$}}
\def\PKS1830{\hbox{PKS\,1830$-$211}}
\def\C2H3+{\hbox{C$_2$H$_3^+$}}
\def\Tform{\hbox{$T_\mathrm{form}$}}

\begin{document}

\title{Protonated acetylene in the z=0.89 molecular absorber toward \PKS1830}

\author{S.~Muller \inst{1}
 \and R.~Le~Gal \inst{2,3}
 \and E.~Roueff \inst{4}
 \and J.\,H.~Black \inst{1}
 \and A. Faure \inst{2}
 \and M.~Gu\'elin \inst{3}
 \and A.~Omont \inst{5}
 \and M.~G\'erin \inst{6}
 \and F.~Combes \inst{7}
 \and S.~Aalto \inst{1}
}
\institute{Department of Space, Earth and Environment, Chalmers University of Technology, Onsala Space Observatory, SE-43992 Onsala, Sweden
  \and Univ. Grenoble Alpes, CNRS, IPAG, 38000 Grenoble, France
  \and Institut de Radioastronomie Millim\'etrique, 300, rue de la piscine, 38406 St Martin d'H\`eres, France
  \and LERMA, Observatoire de Paris, PSL Research University, CNRS, Sorbonne Universit\'e, 92190 Meudon, France
  \and CNRS and Sorbonne Universit\'e, UMR 7095, Institut d’Astrophysique de Paris, 98bis boulevard Arago, 75014 Paris, France
  \and LERMA, Observatoire de Paris, PSL Research University, CNRS, Sorbonne Universit\'e, 75014 Paris, France
  \and Observatoire de Paris, LERMA, College de France, CNRS, PSL Univ., Sorbonne Univ., F-75014, Paris, France
}

\date {Received  / Accepted}

\titlerunning{\C2H3+\ toward \PKS1830}
\authorrunning{S. Muller et al.}

\abstract{We report the first interstellar identification of protonated acetylene, \C2H3+, a fundamental hydrocarbon, in the $z=0.89$ molecular absorber toward the gravitationally lensed quasar \PKS1830. The molecular species is identified from clear absorption features corresponding to the $2_{12}-1_{01}$ (rest frequency 494.034~GHz) and $1_{11}-0_{00}$ (431.316~GHz) ground-state transitions of ortho and para forms of \C2H3+, respectively, in ALMA spectra toward the southwestern image of \PKS1830, where numerous molecules, including other hydrocarbons, have already been detected. From the simple assumption of local thermodynamic equilibrium (LTE) with cosmic microwave background photons and an ortho-to-para ratio of three, we estimate a total \C2H3+\ column density of $2 \times 10^{12}$~cm$^{-2}$ and an abundance of 10$^{-10}$ compared to H$_2$. However, formation pumping could affect the population of metastable states, yielding a \C2H3+\ column density higher than the LTE value by a factor of a few. We explore possible routes to the formation of \C2H3+, mainly connected to acetylene and methane, and find that the methane route is more likely in PDR environment. As one of the initial hydrocarbon building blocks, \C2H3+\ is thought to play an important role in astrochemistry, in particular in the formation of more complex organic molecules.
}
  
\maketitle

\section{Introduction}

There are now more than 300 molecules detected in space (see, e.g., the webpages of the Astrochymist\footnote{http://astrochymist.org/} and the Cologne Database for Molecular Spectroscopy (CDMS\footnote{https://cdms.astro.uni-koeln.de/classic/molecules}), also \citealt{mcgui22}). A few iconic objects have proven to be very prolific sources for detecting new interstellar molecules: Sgr\,B2, a giant molecular cloud close to the Galactic Center (see, e.g., \citealt{num00,bel13,nei14}), TMC\,1, a cold dark cloud in the Taurus complex (see, e.g., \citealt{kai04}, and \citealt{cer21,cer22a,cer22b,cer23,fue22} for recent detections of some hydrocarbon species), L483, a dense core around a class 0 protostar \citep{agundez19,agundez23}, and, particularly, the circumstellar envelope around the evolved star CW~Leo (also known as IRC+10216, see e.g., \citealt{cer00,ten10,par22}).

Concerning extragalactic molecules, two objects stand out in molecule count: the nearby starburst galaxy NGC\,253 (e.g., \citealt{mar06,mar21}), and the $z=0.89$ molecular absorber toward the lensed quasar \PKS1830, for which more than 60 species have been reported to date (\citealt{mul06,mul11,mul14a,ter20}). The latter presents a pure-absorption spectrum, where low-energy transitions are highly favored. As a result, the spectral line density is significantly lower than in rich molecular-emitting sources in closer proximity, which drastically limits line confusion and simplifies line identification. The $z=0.89$ molecular absorber toward \PKS1830\ provides us with a unique opportunity to investigate the interstellar medium and its chemistry in the disk of a distant galaxy (e.g., \citealt{mul17}). The absorber is a nearly face-on ($i=26^\circ$, \citealt{com21}) spiral galaxy which acts as a gravitational lens, splitting the background quasar into two bright and compact images, separated by 1$''$ (\citealt{sub90,jau91,win02}). Molecular absorption is seen against both lensed images \citep{wik98,mul06,mul14a}. The line of sight to the southwestern (SW) image, notably, is the place of the numerous molecular detections. It intercepts the absorber's disk at a projected galactocentric radius of 2.4\,kpc, and has a large column density of H$_2$, $\sim 2 \times 10^{22}$\,cm$^{-2}$, about one order of magnitude larger than the northeastern (NE) line of sight \citep{mul14a}.

Here, we report the identification of a new interstellar molecular ion, protonated acetylene (\C2H3+), in the absorber toward \PKS1830 SW. \C2H3+\ is the protonated form of acetylene (C$_2$H$_2$), the simplest alkyne, i.e., an hydrocarbon with one carbon-carbon triple bond. Due to its symmetry, acetylene does not have permitted dipole rotational transitions and it has to be observed through its infrared vibration-rotation spectrum. Within the Solar system, acetylene is found in planetary atmospheres (e.g., \citealt{rid74,gil75}) and in comets (e.g., \citealt{bro96}). Beyond, acetylene was first detected in IRC+10216 \citep{rid76} and in infrared sources embedded in molecular clouds \citep{lac89}. Its observations tend to be limited to warm environments associated with young stellar objects (e.g., \citealt{bon03,son07,ran18}). The protonated form of acetylene has also been suspected to be present in the interstellar medium as well (e.g., \citealt{her77,gla92}), although it has eluded detection so far, probably due to its unfavorable millimeter-wave rotational spectrum for ground-based observations in local ($z \sim 0$) sources. Both acetylene and its protonated form are key building blocks of larger hydrocarbons and other complex molecules (e.g., \citealt{sch79,oka03,cha09}).

\section{Observations}

\subsection{Ortho-\C2H3+}

We obtained ALMA band~6 observations of \PKS1830\ on 2022 October 11th. The array was composed of 44~antennas in the approximative configuration C-3, with projected baselines ranging from 15 to 500~m resulting in a synthesized beam of $\sim 0.7''$. The antenna primary beam (field of view $\sim 20$'') was much larger than the separation between the lensed images of \PKS1830. The amount of precipitable water vapor was between 0.4 and 0.5~mm during the observations. The total observing time on \PKS1830\ was 36~min. Four spectral windows of 1.875~GHz-wide were used to cover various spectral lines. The spectral window of interest for this discovery was centered at 261.6~GHz and yields a velocity resolution of about 1.1~\kms. It also contained the strong [\ion{C}{i}] ($^3P_1$$-$$^3P_0$) line, redshifted to 260.98~GHz.

The bright quasar J\,1924$-$2914 ($\sim 6.7$~Jy at 260~GHz, at time of the observations) was used as bandpass and flux calibrator. The quasar J\,1832$-$2039 ($\sim 0.3$~Jy) was used for gain calibration. After the standard calibration by the ALMA pipeline (CASA version 6.4.1.12, \citealt{CASA2022}), the visibilities of \PKS1830\ were self-calibrated, phase only, at the shortest integration time interval of 6~sec.

The final spectra along the NE and SW images of \PKS1830\ were extracted by visibility fitting using the Python-based tool {\tt UVMultiFit} \citep{mar14} and a model of two point sources, for which the positions were fixed at the continuum locations and the amplitude per spectral channel left as free parameters. The spectra were then normalized to their corresponding continuum level (i.e., 1.3~Jy and 1.0~Jy at 250~GHz for the NE and SW images, respectively). The rms sensitivity was 0.1\% and 0.08\% of the SW and NE continuum levels, respectively.

\subsection{Para-\C2H3+}

The data described above are complemented by ALMA archival data observed on 2014 March 8th. For these, the array was composed of 25 antennas, with projected baselines ranging between 15 and 420~m, resulting in a synthesized beam $\sim 0.7''$. The amount of precipitable water vapor was $\sim 2$~mm. The total observing time on \PKS1830\ was 1.5~h. The spectral window of interest here was 1.875~GHz wide, centered at 228.5~GHz, and was smoothed to a final velocity resolution of about 1.9~\kms. The calibration was done using J\,1733$-$1304 for the bandpass response of the antennas, Titan for the flux scale, and J\,1923$-$210 for the antenna gains (CASA version 4.2). After self-calibration, the final spectra were extracted the same way as previously explained. The rms noise level was 0.3\% and 0.2\% for the SW and NE normalized spectra, respectively -- at that epoch, the flux density of the NE and SW images at 230~GHz was $\sim 0.5$~Jy and 0.3~Jy, respectively.

\begin{figure}[h] \begin{center}
\includegraphics[width=8.8cm]{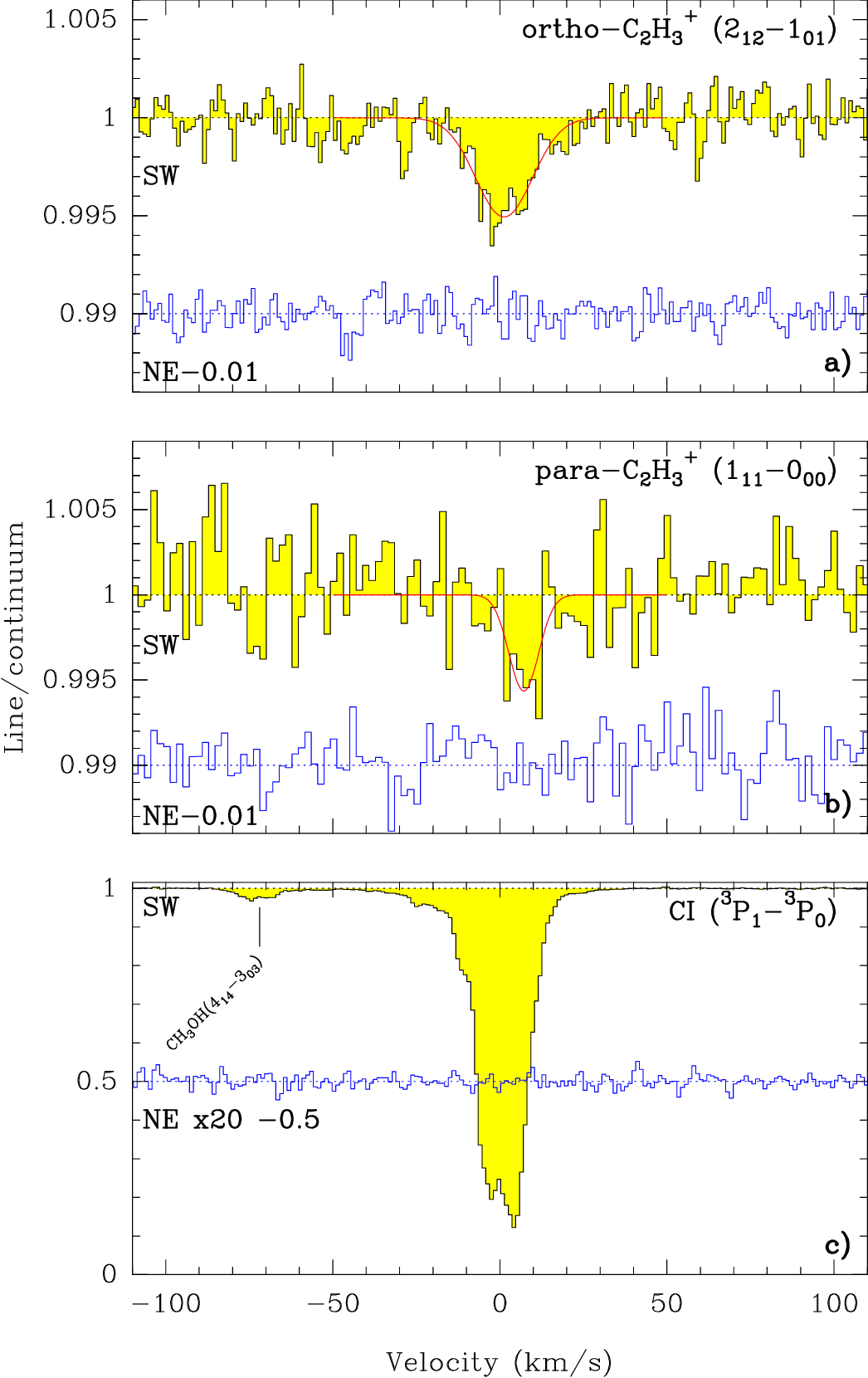}
\caption{ALMA spectra toward the SW and NE lines of sight to \PKS1830: a) The $2_{12}-1_{01}$ transition of ortho-\C2H3+\ (rest frequency 494.034~GHz) observed in 2022. b) The $1_{11}-0_{00}$ transition of para-\C2H3+\ (431.315~GHz, rest), observed in 2014. c) For comparison, spectra of [\ion{C}{i}] ($^3P_1$$-$$^3P_0$) taken from the same spectral window as observations of a). The best fits by a Gaussian line profile are indicated in red on top of the \C2H3+\ lines. All spectra have been normalized to the continuum level of each lensed image. For clarity, the spectra along the NE line of sight have been shifted by $-0.01$ (except for [\ion{C}{i}] which has optical depths multiplied by 20 and shifted by $-0.5$) and are shown in blue.}
\label{fig:spec}
\end{center} \end{figure}

\section{Results and discussion}

\subsection{Identification of \C2H3+}

We started with the serendipitous detection of the absorption feature in Fig.\,\ref{fig:spec}a, which at first we could not identify with any other known species toward \PKS1830. The feature is clear and seen only on the SW line of sight, where many other molecules have already been detected. There is no counterpart in the NE spectra at the same velocities, removing the possibility of spurious instrumental artifacts or of atmospheric lines. Because the two lensed images of \PKS1830\ are only separated by 1\,'', the non-detection of signal on the NE spectra also indicates that the SW absorption feature does not arise from interstellar medium in the Milky Way. We thus have strong evidence that the lines arise in the $z=0.89$ molecular absorber.

Given the shape and signal-to-noise ratio of the line in Fig.\,\ref{fig:spec}a, we adopt a simple Gaussian profile to obtain a first estimate of its rest frequency, $494032 \pm 1$\,MHz, assuming that the line profile is centered at $z=0.88582$ in the heliocentric velocity frame. The total integrated opacity is $(109 \pm 8) \times 10^{-3}$~\kms\ and the full-width at half maximum (FWHM) is $20 \pm 2$~\kms. The linewidth is comparable to previous observations (e.g., \citealt{mul11,mul14a}) and to the FWHM of the [\ion{C}{i}] absorption in Fig.\,\ref{fig:spec}c. Armed with this rest frequency estimate, we have searched for its line identification in the CDMS \citep{CDMS05,CDMS16} and JPL \citep{pic98} molecular databases. As far as we can find exploring a range of $\pm 10$~MHz around 494.032\,GHz (i.e., a range slightly larger than the FWHM), there is only one transition with low energy: the $2_{12}-1_{01}$  ground-state transition of \C2H3+ at rest frequency $494034.0653 \pm 0.0633$~MHz as reported in CDMS. Within the considered frequency range, the next transitions in increasing energy level already have $E_{\rm low} > 80$~K, correspond to more complex molecules, and are unlikely to be detectable at all. For completeness, we also searched around 261.961~GHz (i.e., case where the absorption would occur at $z=0$), and could not find any convincing candidate. The $2_{12}-1_{01}$ transition of \C2H3+\ thus appears as our unique alternative. By repeating the Gaussian fit with this new frequency, we find a velocity centroid $v_0=1.26 \pm 0.68$~\kms\ at the adopted redshift.

At this stage, we already consider \C2H3+\ as the most likely molecule responsible for the absorption feature observed in the SW line of sight of \PKS1830. First, \C2H3+\ is a relatively simple closed-shell molecule with an electric dipole moment of 1.14~D \citep{lee86}, that has been suggested to be present in the interstellar medium (e.g., \citealt{her77,gla92}). Considering the molecular zoo toward \PKS1830, with more than 60 species already detected along the SW line of sight (e.g., \citealt{mul11,mul14a,ter20}), the presence of \C2H3+\ indeed appears highly plausible. In particular, many other hydrocarbons have already been found in the absorber: CH, CH$^+$, C$_2$H, l-C$_3$H, l-C$_3$H$^+$, c-C$_3$H$_2$, l-C$_3$H$_2$, C$_4$H, together with some other complex molecules such as HC$_3$N, CH$_3$CN, CH$_3$CCH, and CH$_2$CHCN. Next, the rest frequency of the transition matches well our best fit estimates within uncertainties. Moreover, the absorption technique favors observations of low-energy level transitions. Especially in the low-density medium (diffuse to translucent clouds), the molecular excitation is easily coupled with photons from the cosmic microwave background, with $\Tcmb = 5.14$~K at $z=0.89$ (see \citealt{mul13}). Indeed, the millimeter-wave absorption spectrum toward \PKS1830\ is completely dominated by ground-state or low-energy transitions. The fact that the transition at 494.034~GHz of \C2H3+\ is directly connected to the ground-state energy level of the more abundant ortho form is therefore a strong asset for its identification. Finally, molecular databases such as the CDMS or JPL are fairly complete which leaves very little chance that they miss any detectable molecule in our distant absorber.

Taking the spectroscopy information from the CDMS (see Sect.\,\ref{sec:spectro}), we list the transitions absorbed from the first levels of \C2H3+\ in Table\,\ref{tab:line} where ortho and para forms are considered separately. We have searched in the ALMA archives and found observations of \PKS1830\ from March 2014 covering the $1_{11}-0_{00}$ line of para-\C2H3+, as shown in Fig.\,\ref{fig:spec}b. The best fit values with a Gaussian profile are the following: $v_0 = 7.2 \pm 0.8$~\kms, ${\rm FWHM} = 10.5 \pm 2.9$~\kms, and an integrated opacity of $(6.0 \pm 1.1) \times 10^{-2}$~\kms. The spectra of the ortho and para lines are not exactly aligned in velocity centroids but clearly overlap in velocity. However, the system is known to have significant time variations of the absorption profiles which are due to structural changes in the background quasar \citep{mul06,mul20,mul21,mul23}. Since the spectra have been taken eight years apart, it is therefore possible that the quasar illumination changed between the two epochs.

Assuming local thermodynamic equilibrium (LTE) conditions, excitation at \Tcmb, and the line parameters as obtained from our Gaussian fittings above, we estimate column densities of $1.4 \times 10^{12}$~cm$^{-2}$ and $0.7 \times 10^{12}$~cm$^{-2}$ for the ortho and para forms of \C2H3+, respectively. Since the lines were observed with a large time interval, we cannot directly sum the column densities of the two forms to get the total column density of \C2H3+, neither measure the ortho-to-para ratio (OPR). However, the values are comparable. From our measurement of ortho-\C2H3+, we derive a total column density of $\sim 2 \times 10^{12}$~cm$^{-2}$ with an assumed ortho-to-para ratio (OPR) of three. This corresponds to an abundance of $10^{-10}$ relative to H$_2$. \C2H3+\ is then $\sim 600$ and 60 times less abundant than C$_2$H and C$_4$H, respectively \citep{mul11}. We note, however, that those column densities could be affected by our simple assumption of LTE. More advanced considerations including formation pumping are discussed in Sect.~\ref{sec:excitation}.
 
Why did \C2H3+\ elude detection at $z=0$ so far? First, the frequencies of the ground-state transitions are difficult or impossible to observe from the ground. The $1_{10}-1_{01}$ transition at 368~GHz falls in a deep atmospheric line. The $1_{11}-0_{00}$ (431~GHz) and $2_{12}-1_{01}$ (494~GHz) transitions are better placed and are accessible in ALMA band~8. We recall that the HIFI receiver of the Herschel Space Observatory was operating above 480~GHz. However, the greatest challenge for the detection of \C2H3+\ is most likely the required sensitivity since the deepest absorption lines toward \PKS1830\ only reach an opacity of $\sim 0.005$ at LTE. In fact, our detection with ALMA could only be achieved because the quasar was in a phase of high flux density, almost a factor two above average measurements done over the last decade (see Fig.\,2 by \citealt{mul23}). Given those challenges, it will be interesting to check whether \C2H3+\ can be detected in other targets, including Galactic sources.

\begin{table*}[ht]
\caption{Line properties for the low-energy level transitions of \C2H3+. Spectroscopic parameters are taken from the CDMS database \citep{CDMS05,CDMS16}} \label{tab:line}
\begin{center} \begin{tabular}{ccrccccccc}

\hline \hline
\multicolumn{2}{c}{Transition} & \multicolumn{1}{c}{Rest Frequency} & Redshifted freq. & $S_{ul}$ & $E_{\rm low}$ & $\tau_{\rm LTE}$ $^{(b)}$ & $\alpha_{\rm LTE}$ $^{(c)}$ & $\tau_{\rm nLTE}$ $^{(d)}$ & \\
         &   &  \multicolumn{1}{c}{ (MHz)}     & $^{(a)}$ (GHz)            &         & (K)         &    ($10^{-3}$)    & ($10^{12}$ cm$^{-2}$\,km$^{-1}$\,s) & ($10^{-3}$)   & \\
\hline
$1_{10}-1_{01}$ & (ortho) & 368572.160 (0.02) & 195.444 & 4.50 & 3.1$^{(e)}$  & 4.5 & 14.8 & 5.0 & \\
$2_{12}-1_{01}$ & (ortho) & 494034.065 (0.06) & 261.973 & 4.50 & 3.1$^{(e)}$  & 4.6 & 14.5 & 5.1 & $^{(f)}$ \\
$2_{12}-3_{03}$ & (ortho) & 166076.743 (0.10) & 88.066 & 3.0 & 18.9 &   0.12 & 572 & 0.60 & \\
$3_{12}-3_{03}$ & (ortho) & 375749.247 (0.02) & 199.250 & 10.4 & 18.9 &   0.48 & 137 & 2.1 & \\
$4_{14}-3_{03}$ & (ortho) & 615171.173 (0.17) & 326.209 &  7.5 & 18.9 &   0.36 & 183 & 1.6 & \\
$4_{14}-5_{05}$ & (ortho)  &  25218.734 (0.04) &  13.373 & 6.3 & 47.2 & \dots & \dots & 4.4 & \\
$5_{14}-5_{05}$ & (ortho)  & 388929.620 (0.03) & 206.239 & 16.1 & 47.2 & \dots & \dots & 11.0 & \\
$6_{16}-5_{05}$ & (ortho)  & 730775.644 (0.34) & 387.511 & 10.7 & 47.2 & \dots & \dots & 7.6 & \\
$7_{16}-7_{07}$ & (ortho)  & 408557.779 (0.03) & 216.647 & 21.3 & 88.1 & \dots & \dots & 1.5 & \\
$8_{18}-7_{07}$ & (ortho)  & 841268.511 (0.56) & 446.102 & 14.1 & 88.1 & \dots & \dots & 1.0 & \\
\hline
$1_{11}-0_{00}$ & (para)  & 431315.570 (0.03) &    228.715 &     1.0 & 0.0 & 1.8 & 11.9 & 1.9 & $^{(f)}$ \\
$1_{11}-2_{02}$ & (para)  & 234505.803 (0.04) &    124.352 &     0.5 &  9.4 &   0.13 & 164 & 0.29 & \\
$2_{11}-2_{02}$ & (para)  & 371430.433 (0.03) &    196.960 &     2.5 &  9.4 &   0.72 & 30.5 & 1.5 & \\
$3_{13}-2_{02}$ & (para)  & 555312.451 (0.11) &    294.467 &     2.0 &  9.4 &   0.60 & 37.0 & 1.2 & \\
$4_{13}-4_{04}$ & (para)   & 381565.745 (0.03) & 202.334 & 4.4 & 31.5 & \dots & \dots & 1.4 & \\
$6_{06}-5_{15}$ & (para)   &  47086.012 (0.03) &  24.968 & 2.6 & 63.8 & \dots & \dots & $-1.0$ & \\
$6_{15}-6_{06}$ & (para)   & 397902.529 (0.03) & 210.997 & 6.3 & 66.1 & \dots & \dots & 2.4 & \\
$7_{17}-6_{06}$ & (para)   & 786626.662 (0.44) & 417.127 & 4.1 & 66.1 & \dots & \dots & 1.6 & \\
\hline
\end{tabular} \end{center}
\tablefoot{ 
(a) Redshifted frequency in the heliocentric rest frame, where the redshift $z=0.88582$ is set.
(b) We assume a total column density N$_{\rm col}({\rm O})+{\rm N}_{\rm col}({\rm P}) = 2 \times 10^{12}$~cm$^{-2}$, an ortho-to-para ratio ${\rm OPR}=3$, and a Gaussian profile with a FWHM linewidth of 20~\kms\ to calculate the line peak opacity $\tau$ under LTE.
(c) The coefficient $\alpha_{\rm LTE}$ is defined as $N_{\rm col}({\rm O,P}) = \alpha_{\rm LTE} \times \int \tau dv$.
(d) Example of non-LTE excitation with guessed collision rates at $T_k = 100$ K, $n({\rm H}_2)=1000$~cm$^{-3}$, $T_{\rm form}=300$~K, destruction rate $9.5\times 10^{-7}$~s$^{-1}$, 
total column density $7.5\times 10^{12}$~cm$^{-2}$, FWHM$ = 20$~\kms. 
(e) The ground-state level of ortho-\C2H3+\ is 3.1~K higher than that of the para form (see Fig.\,\ref{fig:EnergyLevelDiagram}).
(f) Detected in this work.
}
\end{table*}

\subsection{Spectroscopy of \C2H3+} \label{sec:spectro}

Protonated acetylene offers an interesting study case for quantum chemists and spectroscopists alike. The structure of the molecule has long been controversial, but it is now well established that the non-classical planar-bridged (cyclic) structure is more stable than the 'Y'-shaped classical structure \citep{cro89}. Both structures are illustrated in Fig.\,\ref{fig:structure}.

In the bridged form, the molecule can be approximated to first order by a rigid asymmetric rotor (described by the quantum numbers $J$, $K_a$, $K_c$) with two equivalent protons. Transitions are allowed between levels with $\Delta J = 0, \pm 1$, $\Delta K_a=\pm1$, and $\Delta K_c=\pm1$, and can be labeled ortho (between even-odd \and odd-even levels for $K_a$,$K_c$) or para (even-even and odd-odd levels), with statistical weight 3:1, respectively. The energy level diagram of \C2H3+\ is shown in Fig.\,\ref{fig:EnergyLevelDiagram}.

The internal dynamics of the molecule is however complicated by the possible tunneling migration of the protons, interconverting the structure from non-classical to classical with a barrier of about 1600~cm$^{-1}$, as discussed by \cite{hou87}. When considering this tunneling migration, the three protons become equivalent and the even-odd and odd-even levels turn into doublets with a statistical weight of 2:1 with respect to the even-even and odd-odd levels.

Laboratory measurements of the millimeter and submillimeter spectrum of \C2H3+\ were carried out by \cite{bog92} between 195 to 470~GHz. They did not observe line tunneling splitting larger than $\sim 1$~MHz. New measurements were done by \cite{cor96} with improved experimental resolution, where they obtain splittings of the order of a few 100s of kHz, confirming the tunneling. In particular, the $1_{10}-1_{01}$ transition has a splitting of almost 500~kHz. Toward \PKS1830, the splitting is compressed by the factor $(1+z)=1.89$, and it is much smaller than the absorption linewidth. We thus consider it is negligible in the interpretation of our data and we can safely use the CDMS predictions (see Table\,\ref{tab:line}) where these additional splittings are neglected.

\subsection{Excitation of \C2H3+} \label{sec:excitation}

It is interesting to note that, as shown on Fig.\,\ref{fig:EnergyLevelDiagram}, several low lying levels are metastable. They lack dipole-allowed transitions to any lower states. This implies that they could potentially sustain anomalously large populations and associated lines might be brighter than expected from simple LTE estimates.

Even if the chemical source of \C2H3+ remains uncertain, its rate of destruction by dissociative recombination can be estimated, $D = n({\rm e}) k_{DR}$, where 
\begin{equation}
k_{DR}=1.26 \times 10^{-6} (T/100)^{-0.84}~{\rm cm}^3\,{\rm s}^{-1}.
\end{equation}
In steady state, $n(\C2H3+) D$ can be equated with a total formation rate $F$ (cm$^3$\,s$^{-1}$). If we further assume that the formation process populates nascent states of \C2H3+\ with a Boltzmann distribution at some formation temperature \Tform, then \Tform\ is a parameter that governs the OPR, provided that the destruction rate can be assumed to be the same for all states. In this way, source and sink terms can be incorporated into the system of rate equations that controls the populations of all rotational levels \citep{vdtak07}. Crude estimates of inelastic collision rates have been made with a characteristic downward rate coefficient of $2 \times 10^{-10}$~cm$^3$\,s$^{-1}$ for H$_2$ collisions. Inelastic collisions with electrons should also play a role but are neglected for simplicity.
In any case, the details are not very important because the populations of the most important states are controlled by the formation process and by the radiative transitions in the 5.14~K black-body radiation field at $z=0.89$. For example, at $T=100$~K, $n({\rm H}_2)=1000$~cm$^{-3}$, and $n({\rm e}^-)=0.75$~cm$^{-3}$ with a formation temperature of $\Tform=300$~K, we find that the metastable $3_{03}$, $4_{04}$, and $5_{05}$ states have anomalously high populations. Indeed, the $5_{05}$ state is the most highly populated state of all in these conditions, and is depopulated most rapidly by absorption of CMB radiation in the $5_{14}-5_{05}$ and $6_{16}-5_{05}$ transitions at 388.9 and 730.8~GHz, respectively. These transitions are calculated to have higher optical depths than the $2_{12}-1_{01}$ transition at 494~GHz observed by us. Because the metastable states add significantly to the total density, the total column density in this calculation needs to be $7.5 \times 10^{12}$~cm$^{-2}$ (rather than $2 \times 10^{12}$ cm$^{-2}$ in LTE at 5.14~K) in order to produce the observed integrated optical depth at 494~GHz. Moreover, the $4_{14}-5_{05}$ line at 25.2~GHz appears to be refrigerated well below the CMB temperature in this calculation. Given the adopted destruction rate and formation temperature, the OPR is calculated to be 3.125 --- it is not assumed.

In short, although we do not know in detail the state-specific chemical processes or state-to-state collisional rates for \C2H3+, it is likely that its formation and destruction processes play an important role in its excitation, with the possibility that excited metastable states will yield detectable absorption lines.

\subsection{Chemistry of \C2H3+}

We here discuss the possible gas-phase formation channels of \C2H3+, which is observed with an abundance $\simeq 10^{-10}$ relative to H$_2$ toward \PKS1830 SW. First, C$_2$H$_3^+$ does not react with H$_2$ and is rapidly destroyed by dissociative recombination with electrons via the two dominant channels 1,2 in Table~\ref{tab:chemistry} (with branching ratios of 59\% and 29\%, respectively, \citealt{kal02}).
The total destruction rate coefficient is $k_e \sim (9 - 1) \times 10^{-6}$~cm$^3$\,s$^{-1}$ at 10--100~K, respectively. Assuming steady-state kinetics and that the main formation route to \C2H3+\ involves parent species A$^+$ and B (A$^+$ + B $\longrightarrow$ \C2H3+\ + products), with a typical formation rate $k_f \sim 10^{-9}$~cm$^3$\,s$^{-1}$, we then can write:
\begin{equation} \label{eq:AB}
[\C2H3+] = \frac{k_f}{k_e} \frac{[{\rm A}^+][{\rm B}]}{[{\rm e}^-]}.
\end{equation}
Since $[{\rm A}^+] < [{\rm e}^-]$, we get $[\C2H3+] < [{\rm B}]/100$.
In \PKS1830 SW, we would thus require $[{\rm B}]> 10^{-8}$ relative to H$_2$.

Hereafter, we discuss the cases where B=C$_2$H$_2$ and CH$_4$, from the reaction list in Table~\ref{tab:chemistry}.

\subsubsection{The route of acetylene}

Both acetylene and its protonated ion are key species involved in the rich hydrocarbon chemistry of the interstellar medium (e.g., \citealt{sch79,her89,oka03}). 

Protonated forms MH$^+$ of symmetric molecules M without dipole moment are interesting observational proxies of these unobservable species at radio wavelengths (e.g., \citealt{her77}), provided M and MH$^+$ are clearly chemically connected (i.e., proton exchange reaction: M + XH$^+$ $\longrightarrow$ MH$^+$ + X). 

Proton donors to acetylene can be, e.g., the H$_3^+$, HCO$^+$, and N$_2$H$^+$ ions (Reactions 4,5,6 from Table~\ref{tab:chemistry}), with typical rates $\sim 10^{-9}$~cm$^3$\,s$^{-1}$. Adopting Eq.\,\ref{eq:AB} for example with B=HCO$^+$ (reaction 5 in Table~\ref{tab:chemistry}), we would get:
\begin{equation}
\frac{[{\rm C}_2{\rm H}_2]}{[\C2H3+]} = 7 \times 10^3 - 7 \times 10^7
\end{equation}
with [HCO$^+$]/[H$_2$] $=9\times 10^{-9}$ (as measured toward \PKS1830 SW, \citealt{mul11,mul14a}) and [e$^-$]/[H$_2$] $= 10^{-8} - 10^{-4}$, respectively, depending on the environment.

Testing such simple chemical considerations with more elaborated gas-phase models of dark-cloud chemistry, \cite{agu15} found a good empirical correlation between the abundance ratio of a molecule M and its protonated form MH$^+$ and the proton affinity of M. Considering the proton affinity of acetylene ($\sim 150$~kcal/mol), their correlation would yield a ratio [C$_2$H$_2$]/[\C2H3+] $\sim 10^4$. \cite{agu22} updated their diagram of [MH$^+$]/[M] abundance ratio versus proton affinity with observations of new protonated-neutral pairs and confirmed the trend. Basically, they argue that if the proton affinity of molecule M is lower than that of CO, then the main proton donor forming MH$^+$ would be H$_3^+$. If the proton affinity is higher, then the main donor would be HCO$^+$. Since the later is relatively abundant in cold dense clouds, the formation of MH$^+$ could then be enhanced by orders of magnitude, moving the [M]/[MH$^+$] ratio from a typical range $10^4 - 10^6$ down to values of $10 - 10^3$.

It is worth noticing that the [MH$^+$]/[M] ratio is in principle sensitive to the degree of ionization, and thus, in particular to the cosmic-ray ionization rate, $\zeta$. From OH$^+$ and H$_2$O$^+$ absorption, \cite{mul16} did estimate a value $\zeta \sim 2 \times 10^{-14}$~s$^{-1}$ in the SW line of sight through the absorber. This value is comparable to values derived in our Galactic Center, but significantly higher (by about 1-2 orders of magnitude) than in the Galactic disk. The abundance of \C2H3+\ in the $z=0.89$ absorber (SW line of sight) could then be enhanced by the elevated $\zeta$.

Perhaps l-C$_3$H$^+$, recently identified toward \PKS1830 by \cite{ter20}, is a better indirect tracer of acetylene than \C2H3+, if its main source is ${\rm C}^+ + {\rm C}_2{\rm H}_2 \longrightarrow {\rm C}_3{\rm H}^+ + {\rm H}$. Although l-C$_3$H$^+$ reacts with H$_2$, the rate coefficient is low, $5.2 \times 10^{-12}$~cm$^3$\,s$^{-1}$, so that it is likely to be destroyed mainly by dissociative recombination. In that case, and if PDR conditions prevail, $[{\rm C}^+]\approx [{\rm e}^-]$, then [C$_2$H$_2$]/[l-C$_3$H$^+$]~$\approx 200$ at $T=100$~K. Given the column density found by \cite{ter20}, N(C$_3$H$^+$)~$=1.2 \times 10^{12}$~cm$^{-2}$, the implied abundance of acetylene is $\sim 10^{-8}$ relative to H$_2$. If this is correct, then the abundance of C$_2$H$_2$ in \PKS1830 SW is likely too small for it to be the main parent of the observed \C2H3+.

\subsubsection{The route of methane}

Other channels to the formation of \C2H3+\ involve reactions of CH$_4$ with C$^+$ or CH$^+$ as parent molecules (reactions 7,8 in Table~\ref{tab:chemistry}). These reactions may indeed be particularly efficient in diffuse and translucent clouds where the abundance of C$^+$ and CH$^+$ is high. Direct observations of interstellar methane were first made by \cite{lac91}, who found CH$_4$/CO~$\approx 10^{-3}$ in the gaseous phase toward several IR-bright protostars. Subsequently \cite{boo98} found CH$_4$/H$_2$~$\approx 4\times 10^{-7}$ in warm, dense gas toward protostars. Most recently, \cite{nic23} analyzed high-resolution SOFIA/EXES mid-infrared spectra of several species including CH$_4$ and C$_2$H$_2$ toward Orion~IRc2. Based on the measurements at 7.6~$\mu$m wavelength, CH$_4$/H$_2$~$\approx 1.5\times 10^{-6}$ and CH$_4$/C$_2$H$_2$~$=1.5$ there. These hot-core or protostellar sources may have such a large CH$_4$ abundance owing to evaporation of ices from dust surfaces. 

In PDR conditions where $n({\rm C}^+) \approx n({\rm e})$ , the C$^+$~+~CH$_4$ source of \C2H3+\ can maintain an abundance [\C2H3+]/[H$_2$]~$= 1\times 10^{-10} ([{\rm CH}_4] / 1.3\times 10^{-7})$. In other words, the estimated abundance of \C2H3+\ toward \PKS1830 SW can be explained by PDR conditions if methane is present at 1/10 of the abundance that it exhibits in hot core regions of the Milky Way like the material toward Orion~IRc2. On the other hand, C$_2$H$_2$ is not a viable source of \C2H3+\ even when its abundance is as high as observed toward Orion~IRc2, because the reactant ions (H$_3^+$ and HCO$^+$) have much lower abundances than C$^+$ in PDR conditions.

\begin{table*}[ht]
\caption{Main chemical reactions involving \C2H3+\ in the interstellar medium (taken from KIDA,  \citealt{KIDA}).} \label{tab:chemistry}
\begin{center} \begin{tabular}{cccccc}
\hline \hline
Type & \multicolumn{3}{c}{Reactions} & $k(T)$ & Label \\
       &   & &        & (cm$^3$\,s$^{-1}$) & \\
\hline
Dissociative recombination   & \C2H3+\ + e$^-$ & $\longrightarrow$ & C$_2$H + H + H   & $k_{1}(100 {\rm K}) =7.4 \times 10^{-7}$ & 1 \\
                            &                 & $\longrightarrow$ & C$_2$H$_2$ + H   & $k_{2}(100 {\rm K}) =3.6 \times 10^{-7}$& 2 \\  

\hline
Destruction by H atom &  \C2H3+\ + H & $\longrightarrow$ & C$_2$H$_2^+$ + H$_2$   & $k_{3} =6.8 \times 10^{-11}$ & 3 \\
\hline
Protonation of acetylene &  C$_2$H$_2$ + H$_3^+$ & $\longrightarrow$ & \C2H3+\ + H$_2$ & $k_4 = 3.5 \times 10^{-9}$ & 4 \\
           &  C$_2$H$_2$ + HCO$^+$ & $\longrightarrow$ & \C2H3+\ + CO & $k_5 = 1.4 \times 10^{-9}$ & 5 \\ 
           &  C$_2$H$_2$ + N$_2$H$^+$ & $\longrightarrow$ & \C2H3+\ + N$_2$ &  $k_6 = 1.4 \times 10^{-9}$ & 6 \\ 
\hline
Production from methane & CH$_4$ + C$^+$ & $\longrightarrow$ & \C2H3+\ + H & $k_7 = 1.0 \times 10^{-9}$ & 7 \\
       & CH$_4$ + CH$^+$ & $\longrightarrow$ & \C2H3+\ + H$_2$ & $k_8 = 1.1 \times 10^{-9}$ & 8 \\
\hline
  & CH$_2$ + CH$_3^+$ & $\longrightarrow$ & \C2H3+\ + H$_2$ &  $k_9(100 {\rm K}) =2.3 \times 10^{-9}$    &  9 \\ 
\hline
Hydrogenation of C$_2$H$_2^+$  & C$_2$H$_2^+$ + H$_2$ & $\longrightarrow$ & \C2H3+\ + H  & $k_{10}\la 3\times 10^{-12}$ at 
$T\leq 300$ K & 10 \\
\hline
Top-down chemistry e.g.: & l-C$_3$H$_3^+$ + O & $\longrightarrow$ & \C2H3+ + CO & $k_{11} = 4.5 \times 10^{-11}$ & 11  \\
\hline
PAH photo-degradation  & PAH + $\gamma$ & $\longrightarrow$ & \C2H3+ + products & -- & 12 \\ 
\hline
\end{tabular} \end{center}
\end{table*}

\subsubsection{Other reactions}
 
For completeness, we briefly list some other reactions involving \C2H3+. First, the slightly endothermic hydrogenation reaction of C$_2$H$_2^+$ (reaction 10 of Table~\ref{tab:chemistry}, see, e.g., discussion by \citealt{oka03} and \citealt{smi93}) may take place in regions where excess energy is present (e.g., being radiative, or due to shock or turbulence dissipation). C$_2$H$_3^+$ may also interact either with other neutrals (HCN, NH$_3$, H$_2$O) to form back acetylene or with simple hydrocarbons to produce more complex hydrocarbon ions \citep{sch79,cha09}, e.g.:
\begin{equation}
{\rm C}_2{\rm H}_3^+ + {\rm CH}_4 \longrightarrow {\rm C}_3{\rm H}_5^+ + {\rm H}_2,
\end{equation}
\begin{equation}
{\rm C}_2{\rm H}_3^+ + {\rm C}_2{\rm H}_2 \longrightarrow {\rm C}_4{\rm H}_3^+ + {\rm H}_2,
\end{equation}
\begin{equation}
{\rm C}_2{\rm H}_3^+ +{\rm HC}_{\rm n+1}{\rm N}  \longrightarrow {\rm H}_2{\rm C}_{\rm n+3}{\rm N}^+ + {\rm H}_2.
\end{equation}

On the other hand, \C2H3+\ could be the product of top-down chemistry, e.g.,
\begin{equation}
{\rm C}_3{\rm H}_3^+ +{\rm O} \longrightarrow {\rm C}_2{\rm H}_3^+ + {\rm CO},
\end{equation}
or even PAH photo-degradation \citep{all96} that produces acetylene, which can further be protonated as previously described.

Let us here finally recall that long carbon chains remain important candidates for the unknown carriers of the diffuse interstellar bands \citep{dou77,cox14}, including perhaps protonated forms such as HC$_{2n}$H$_2^+$.

\subsection{Nature of the \C2H3+\ absorbing gas toward \PKS1830}

There is evidence for a multi-phase medium in the absorbing column toward the SW image of \PKS1830. A component with low  molecular hydrogen fraction (less than or about a few percent) is traced by hydrides such as ArH$^+$ \citep{hmul15}, OH$^+$ and H$_2$O$^+$ \citep{mul16}, CH$^+$ \citep{mul17}, and H$_2$Cl$^+$ \citep{mul14b}, which all have broad absorption profiles. In contrast, absorptions from CH and H$_2^{18}$O \citep{mul23} show a narrower profile, with a higher H$_2$-fraction (10-100\%). More complex molecules like CH$_3$OH, HC$_3$N, CH$_3$NH$_2$ have their line barycenter offset compared to the bulk of other species \citep{mul11}, suggesting further chemical segregation. Finally, the time variability of the absorption profiles, caused by structural changes in the quasar's morphology \citep{mul08}, has highlighted at least one velocity component that shows similarities with Galactic dark clouds, both in physical conditions and chemical composition \citep{mul21,mul23}.

With the limited signal-to-noise ratio of the \C2H3+\ spectrum in Fig.\ref{fig:spec}ab and the time variability of the system, it is difficult to conclude on the nature of the \C2H3+\ absorbing gas by comparing its line profile to previous observations. However, observations of future variations may reveal a correlation between \C2H3+\ and other species, giving us clues on its origin.
 
Finally, we note that the simultaneous detection of both the ortho and para forms would allow us to measure the actual OPR of \C2H3+. This ratio could deviate from the statistical value of three and thus possibly help us to discriminate among the formation processes discussed above (see, e.g., \citealt{fau13} for a discussion on ``anomalous" OPR).

\section{Summary and conclusions}

We report the detection of protonated acetylene (\C2H3+) in ALMA absorption spectra toward the southwestern image of the lensed quasar \PKS1830. This constitutes the first detection of \C2H3+\ in space, quite remarkably in a source at $z=0.89$. Assuming LTE equilibrium with CMB and an OPR of three, we estimate a total \C2H3+\ column density of $2 \times 10^{12}$~cm$^{-2}$, corresponding to an abundance of $10^{-10}$ relative to H$_2$. This makes the species quite challenging to detect, with small line opacities $\lesssim 0.005$ toward \PKS1830 SW. However, we argue that formation pumping could increase the population of metastable states, resulting in more easily detectable absorption lines.

We explore possible gas-phase routes to the formation of interstellar \C2H3+. The ion mainly connects either to acetylene or methane, ironically both symmetric and lacking dipole moment hence unobservable at radio wavelengths. The methane route is shown to be likely more efficient in PDR conditions.

The discovery of \C2H3+\ (and recently of H$_2$NC, \citealt{cab21}) toward \PKS1830\ highlights the potentials of absorption spectroscopy against this quasar to search for, and potentially find other ``missing" interstellar molecules, taking advantage of {\em i)} the greatly reduced line confusion in the absorption spectrum, {\em ii)} the red-shifting of the lines into potentially more favorable frequencies for ground-based observations, and {\em iii)} periods when the quasar is bursting in millimeter waves for augmented line-detection sensitivity.

\begin{acknowledgement}
We thank the referee and language editor for their help improving the clarity of the manuscript.
This paper makes use of the following ALMA data: ADS/JAO.ALMA\#2022.1.01634.S (ortho-\C2H3+ and [\ion{C}{i}]) and ADS/JAO.ALMA\#2012.1.00581.S (para-\C2H3+). ALMA is a partnership of ESO (representing its member states), NSF (USA) and NINS (Japan), together with NRC (Canada), MOST and ASIAA (Taiwan), and KASI (Republic of Korea), in cooperation with the Republic of Chile. The Joint ALMA Observatory is operated by ESO, AUI/NRAO and NAOJ. This research has made use of the NASA's Astrophysics Data System, the Cologne Database for Molecular Spectroscopy (CDMS, https://cdms.astro.uni-koeln.de/classic/molecules, \citealt{CDMS05,CDMS16}), the JPL Molecular Spectroscopy database (https://spec.jpl.nasa.gov, \citealt{pic98}), and of the KInetic Database for Astronomy (KIDA, https://kida.astrochem-tools.org, \citealt{KIDA}).
\end{acknowledgement}

\clearpage

\begin{appendix}

\section{Complementary material}

\begin{figure}[h] \begin{center}
\includegraphics[width=8.cm]{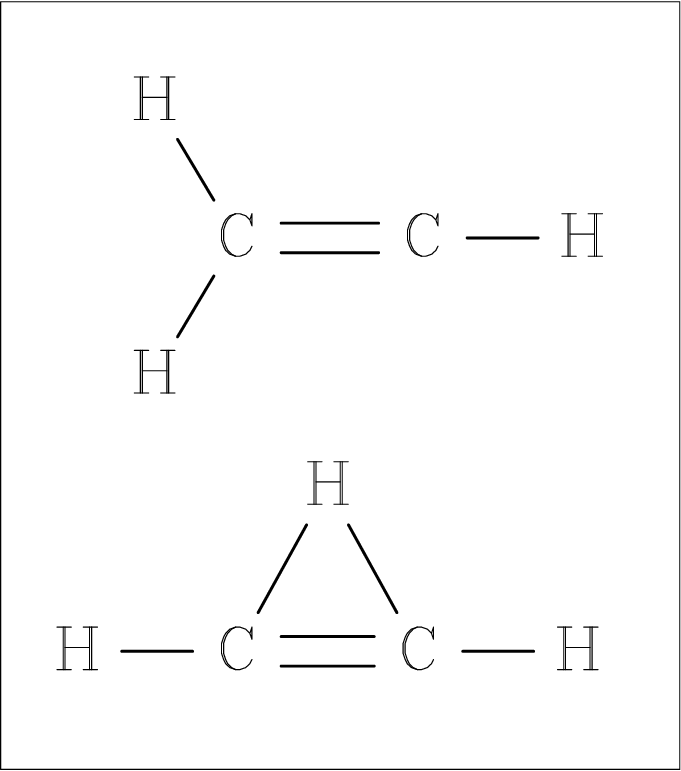}
\caption{Illustration showing the possible structures of \C2H3+. Up: classical 'Y'-shaped. Down: non-classical planar-bridged cyclic. The lab data support the bridged structure over the classical one \citep{cro89,cor96}.}
\label{fig:structure}
\end{center} \end{figure}

\begin{figure*}[h] \begin{center}
\includegraphics[width=\textwidth]{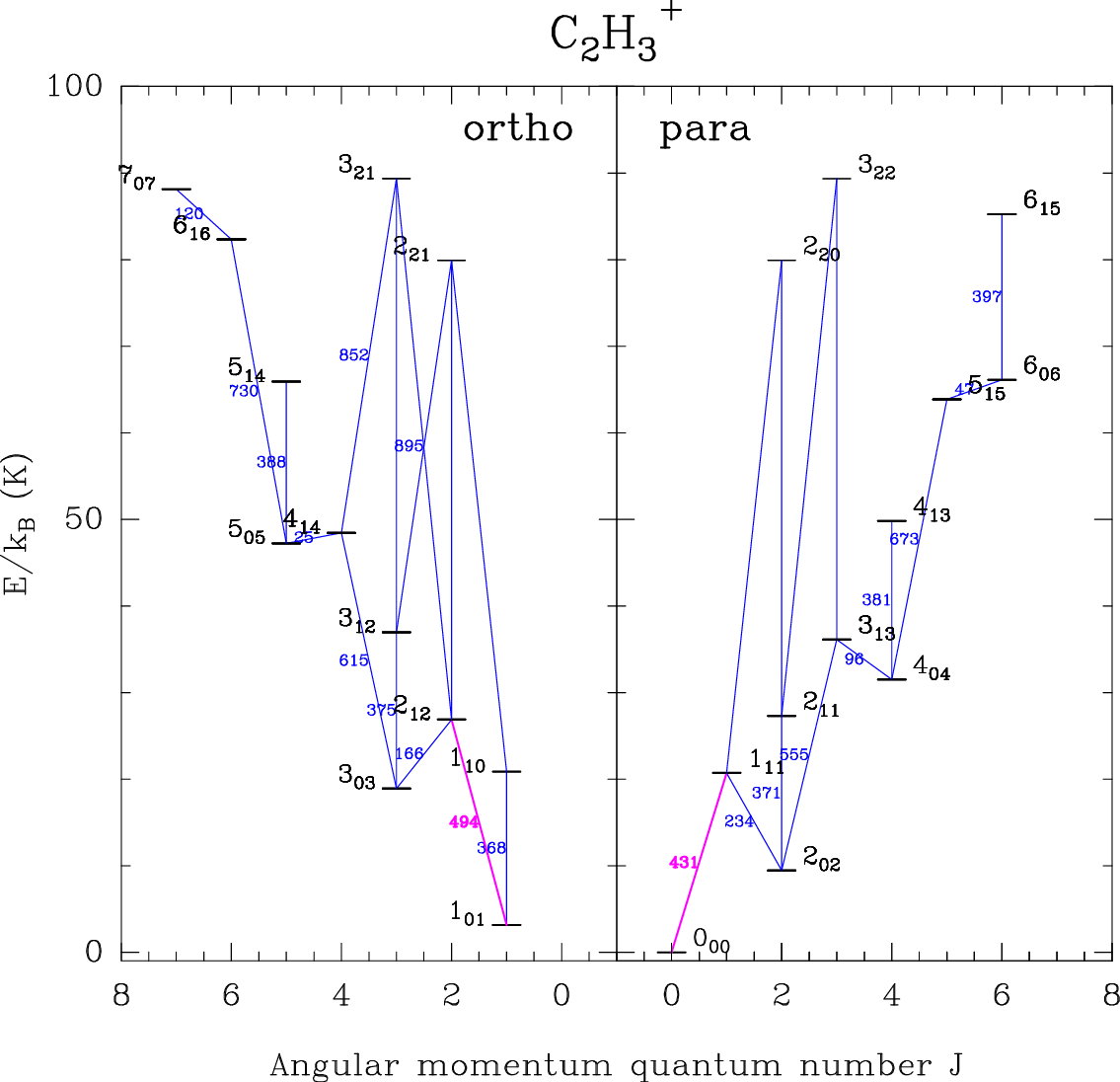}
\caption{Diagram of the energy levels of \C2H3+\ for levels below 100~K. The labels correspond to the $J_{K_a K_c}$ quantum numbers. The blue lines indicate the allowed rotational transitions, with frequencies given (in GHz), if below 1~THz. The lines observed in this work are shown in magenta.}
\label{fig:EnergyLevelDiagram}
\end{center} \end{figure*}

The possible structures of \C2H3+\ are illustrated in Fig.\,\ref{fig:structure}. A diagram of energy levels of \C2H3+\ for levels below 100~K is shown in Fig.\,\ref{fig:EnergyLevelDiagram}.

\end{appendix}


\begin{thebibliography}{}


\bibitem[Ag\'undez et al.(2015)]{agu15}{Ag\'undez, M., Cernicharo, J., de Vicente, P., et al., 2015, \aap, 579, L10}
\bibitem[Ag\'undez et al.(2019)]{agundez19}{Ag\'undez, M., Marcelino, N., Cernicharo, J., Roueff, E., and Tafalla, M., 2019, \aap, 625, A147}
\bibitem[Ag\'undez et al.(2022)]{agu22}{Ag\'undez, M., Cabezas, C., Marcelino, N., et al., 2022, \aap, 659, L9}
\bibitem[Ag\'undez et al.(2023)]{agundez23}{Ag\'undez, M., Marcelino, N., Tercero, B., and Cernicharo, J., 2023, \aap, L13}
\bibitem[Allain et al.(1996)]{all96}{Allain, T., Leach, S., \& Sedlmayr, E., 1996, \aap, 305, 602} 
\bibitem[Belloche et al.(2013)]{bel13}{Belloche, A., M\"uller, H. S. P., Menten, K. M., Schilke, P., \& Comito, C., 2013, \aap, 559, 47}
\bibitem[Bogey et al.(1992)]{bog92}{Bogey, M., Cordonnier, M., Demuynck, C., \& Destombes, J. L., 1992, \apj, 399, 103}
\bibitem[Boogert et al.(1998)]{boo98}{Boogert, A. C. A., Helmich, F. P., van Dishoeck, E. F., et al., 1998, \aap, 336, 352} 
\bibitem[Boonman et al.(2003)]{bon03}{Boonman, A. M. S., van Dishoeck, E. F., Lahuis, F., et al., 2003, \aap, 399, 1047}
\bibitem[Brooke et al.(1996)]{bro96}{Brooke, T. Y., Tokunaga, A. T., Weaver, H. A., et al., 1996, \nat, 383, 606}
\bibitem[Cabezas et al.(2021)]{cab21}{Cabezas, C., Ag\'undez, M., Marcelino, N., et al., 2021, \aap, 654, 45} 
\bibitem[The CASA Team(2022)]{CASA2022}{The CASA Team, et al., “CASA, the Common Astronomy Software Applications for Radio Astronomy”, PASP, 2022, 134, 114501 DOI: 10.1088/1538-3873/ac9642}
\bibitem[Cernicharo et al.(2000)]{cer00}{Cernicharo, J., Gu\'elin, M., \& Kahane, C., 2000, \aaps, 142, 181}
\bibitem[Cernicharo et al.(2021)]{cer21}{Cernicharo, J.; Ag\'undez, M.; Kaiser, R. I., et al., 2021, \aap, 652, 9} 
\bibitem[Cernicharo et al.(2022a)]{cer22a}{Cernicharo, J.; Ag\'undez, M.; Cabezas, C., et al., 2022a, \aap, 657, 16} 
\bibitem[Cernicharo et al.(2022b)]{cer22b}{Cernicharo, J.; Fuentetaja, R.; Ag\'undez, M., 2022b, \aap, 663, 9} 
\bibitem[Cernicharo et al.(2023)]{cer23}{Cernicharo, J.; Tercero, B.; Marcelino, N., 2023, \aap, 674, 4} 
\bibitem[Chapman et al.(2009)]{cha09}{Chapman, J. F., Millar, T. J., Wardle, M., et al., 2009, \mnras, 394, 221}
\bibitem[Combes et al.(2021)]{com21}{Combes, F., Gupta, N., Muller, S., et al., 2021, \aap, 648, 116}
\bibitem[Cordonnier \& Coudert(1996)]{cor96}{Cordonnier, M. \& Coudert, L.~H., 1996, Journal of Molecular Spectroscopy, 178, 59}
\bibitem[Cox \& Cami(2014)]{cox14}{Cox, N.~L.~J., \& Cami, J.\ 2014, IAU Symposium, 297, 412}
\bibitem[Crofton et al.(1989)]{cro89}{Crofton, M. W., Jagod, M.-F., Rehfuss, B. D., \& Oka, T., 1989, J. Chem. Phys. 91, 5139}
\bibitem[Douglas(1977)]{dou77}{Douglas, A.~E.\ 1977, \nat, 269, 130}
\bibitem[Endres et al.(2016)]{CDMS16}{Endres, C. P., Schlemmer, S., Schilke, P., Stutzki, J., \& M\"uller, H. S. P., 2016, Journal of Molecular Spectroscopy, 327, 95}
\bibitem[Faure et al.(2013)]{fau13}{Faure, A., Hily-Blant, P., Le Gal, R., Rist, C., \& Pineau des For\^ets, G., 2013, \apjl, 770, L2}
\bibitem[Fuentetaja et al.(2022)]{fue22}{Fuentetaja, R.; Ag\'undez, M.; Cabezas, C., 2022, \aap, 667, 4} 
\bibitem[Gillett(1975)]{gil75}{Gillett, F. C., 1975, \apj, 201, 41} 
\bibitem[Glassgold, Omont \& Gu\'elin(1992)]{gla92}{Glassgold, A. E., Omont, A., \& Gu\'elin, M., 1992, \apj, 396, 115}
\bibitem[Godard et al.(2014)]{god14}{Godard, B., Falgarone, E., \& Pineau des For{\^e}ts, G.\ 2014, \aap, 570, 27}
\bibitem[Herbst et al.(1977)]{her77}{Herbst, E., Green, S., Thaddeus, P., Klemperer, W., 1977, \apj, 215, 503}
\bibitem[Herbst \& Leung(1989)]{her89}{Herbst, E. \& Leung, C. M., 1989, \apjs, 69, 271}
\bibitem[Hougen(1987)]{hou87}{Hougen, J.T., 1987, Journal of Molecular Spectroscopy, 123, 197}
\bibitem[Jauncey et al.(1991)]{jau91}{Jauncey, D. L., Reynolds, J. E., Tzioumis, A. K., et al., 1991, \nat, 352, 132}
\bibitem[Kaifu et al.(2004)]{kai04}{Kaifu, N., Ohishi, M., Kawaguchi, K., et al., 2004, \pasj, 56, 69} 
\bibitem[Kalhori et al.(2002)]{kal02}{Kalhori, S., Viggiano, A.A., Arnold, S.T., et al., 2002, \aap, 391, 1159} 
\bibitem[Lacy et al.(1989)]{lac89}{Lacy, J. H., Evans, N. J., Achtermann, J. M., et al., 1989, \apjl, 342, L43}
\bibitem[Lacy et al.(1991)]{lac91}{Lacy, J. H., Carr, J. S., Evans, Neal J., II, et al., 1991, \apj, 376, 556} 
\bibitem[Lee \& Schaefer(1986)]{lee86}{Lee T. J. \& Schaefer H. F., 1986, J. Chem. Phys., 85, 3437} 
\bibitem[McGuire(2022)]{mcgui22}{McGuire, B. A., 2022, \apjs, 259, 30}
\bibitem[Mart\'in et al.(2006)]{mar06}{Mart\'in, S., Mauersberger, R., Mart\'in-Pintado, J., et al., 2006, \apjs, 164, 450}
\bibitem[Mart\'in et al.(2021)]{mar21}{Mart\'in, S., Mangum, J. G., Harada, N., et al., 2021, \aap, 656, 46}
\bibitem[Mart\'i-Vidal et al.(2014)]{mar14}{Mart\'i-Vidal, I., Vlemmings, W., Muller, S., \& Casey S., 2014, \aap, 563, 136} 
\bibitem[M{\"u}ller et al.(2005)]{CDMS05}{M{\"u}ller, H.~S.~P., Schl{\"o}der, F., Stutzki, J., \& Winnewisser, G., 2005, J. Mol. Struct., 742, 215}
\bibitem[M\"{u}ller et al.(2015)]{hmul15}{M\"uller, H. S. P., Muller, S., Schilke, P., et al., 2015, \aap, 582, L4} 
\bibitem[Muller et al.(2006)]{mul06}{Muller, S., Gu\'elin, M., Dumke, M., et al., 2006, \aap, 458, 417} 
\bibitem[Muller \& Gu\'elin(2008)]{mul08}{Muller, S. \& Gu\'elin, M., 2008, \aap, 491, 739} 
\bibitem[Muller et al.(2011)]{mul11}{Muller, S., Beelen, A., Gu\'elin, M., et al., 2011, \aap, 535, 103} 
\bibitem[Muller et al.(2013)]{mul13}{Muller, S., Beelen, A., Black, J. H., et al., 2013, \aap, 551, 109} 
\bibitem[Muller et al.(2014a)]{mul14a}{Muller, S., Combes, F., Gu\'elin, M., et al., 2014, \aap, 566, 112} 
\bibitem[Muller et al.(2014b)]{mul14b}{Muller, S., Black, J. H., Gu\'elin, M., et al., 2014b, \aap, 566, 6} 
\bibitem[Muller et al.(2016)]{mul16}{Muller, S., M\"uller, H. S. P., Black, J. H., et al., 2016, \aap, 595, 128} 
\bibitem[Muller et al.(2017)]{mul17}{Muller, S., M\"uller, H. S. P., Black, J. H., et al., 2017, \aap, 606, 109} 
\bibitem[Muller et al.(2020)]{mul20}{Muller, S., Roueff, E., Black, J. H., et al., 2020, \aap, 637, 7} 
\bibitem[Muller et al.(2021)]{mul21}{Muller, S., Ubachs, W., Menten, K. M., et al, 2021, \aap, 652, 5} 
\bibitem[Muller et al.(2023)]{mul23}{Muller, S., Mart\'i-Vidal, I., Combes, F., et al., 2023, \aap, 674, 101}
\bibitem[Neill et al.(2014)]{nei14}{Neill, J. L. Bergin, E. A., Lis, D. C., 2014, \apj, 789, 8}
\bibitem[Nickerson et al.(2023)]{nic23}{Nickerson, S., Rangwala, N.; Colgan, S. W. J., et al., 2023, \apj, 945, 26} 
\bibitem[Nummelin et al.(2000)]{num00}{Nummelin, A., Bergman, P., Hjalmarson, \AA, et al., 2000, \apjs, 128, 213}
\bibitem[Oka et al.(2003)]{oka03}{Oka, T., Thorburn, J. A., McCall, B. J., et al., 2003, \apj, 582, 823} 
\bibitem[Pardo et al.(2022)]{par22}{Pardo, J. R., Cernicharo, J., Tercero, B., 2022, \aap, 658, 39}
\bibitem[Pickett et al.(1998)]{pic98}{Pickett, H. M., Poynter, I. R. L., Cohen, E. A., et al., 1998, \jqsrt, 60, 883} 
\bibitem[Rangwala et al.(2018)]{ran18}{Rangwala, N., Colgan, S. W. J., Le Gal, R., et al., 2018, \apj, 856, 9} 
\bibitem[Ridgway(1974)]{rid74}{Ridgway, S. T., 1974, \apj, 187, 41}
\bibitem[Ridgway et al.(1976)]{rid76}{Ridgway, S. T., Hall, D. N. B., Wojslaw, R. S., et al., 1976, \nat, 264, 345}
\bibitem[Schiff \& Bohme(1979)]{sch79}{Schiff, H. I. \& Bohme, D. K., 1979, \apj, 232, 740}
\bibitem[Smith et al.(1993)]{smi93}{Smith, D., Glosik, J., Skalsky, V., {\v S}pan{\v e}l, P., Lindinger, W. 1993, Int. J. Mass Spectrometry Ion Proc., 129, 145}
\bibitem[Sonnentrucker et al.(2007)]{son07}{Sonnentrucker, P., Gonz\'alez-Alfonso, E., \& Neufeld, D. A., 2007, \apj, 671, L37} 
\bibitem[Subrahmanyan et al.(1990)]{sub90}{Subrahmanyan, R., Narasimha, D., Pramesh-Rao, A., \& Swarup, G., 1990, \mnras, 246, 263}
\bibitem[Tercero et al.(2020)]{ter20}{Tercero, B., Cernicharo, J., Cuadrado, S., de Vicente, P., \& Gu\'elin, M., 2020, \aap, 636, 7}
\bibitem[Tenenbaum et al.(2010)]{ten10}{Tenenbaum, E. D., Dodd, J. L., Milam, S. N., Woolf, N. J., \& Ziurys, L. M., 2010, \apjs, 190, 348}
\bibitem[van der Tak et al.(2007)]{vdtak07}{van der Tak, F. F. S., Black, J. H., Sch\"oier, F. L., Jansen, D. J., van Dishoeck, E. F., 2007, \aap, 468, 627}
\bibitem[Wakelam et al.(2012)]{KIDA}{Wakelam, V., Herbst, E., Loison, J.-C., et al., 2012, \apjs, 199, 21} 
 \bibitem[Wiklind \& Combes(1998)]{wik98}{Wiklind, T. \& Combes, F., 1998, \apj, 500, 129}
\bibitem[Winn et al.(2002)]{win02}{Winn, J. N., Kochanek, C. S., McLeod, B. A., et al., 2002, \apj, 575, 103}

  
\end{thebibliography}
\end{document}